\documentclass[a4paper]{jpconf}
\usepackage{graphicx,amsmath}
\begin{document}
\title{Heat current through an artificial Kondo impurity beyond linear response}

\author{Miguel A. Sierra and David S\'anchez}

\address{IFISC (UIB-CSIC), Campus Universitat Illes Balears, E-07122 Palma de Mallorca, Spain}

\ead{msierra@ifisc.uib-csic.es}

\begin{abstract}
We investigate the heat current of a strongly interacting quantum dot in the presence of a voltage bias in the Kondo regime. Using the slave-boson mean-field theory, we discuss the behavior of the energy flow and the Joule heating.
We find that both contributions to the heat current display interesting symmetry properties under reversal
of the applied dc bias. We show that the symmetries arise from the behavior of the dot transmission function.
Importantly, the transmission probability is a function of both energy and voltage. This allows us to analyze
the heat current in the nonlinear regime of transport. We observe that nonlinearities appear already
for voltages smaller than the Kondo temperature. Finally, we suggest to use the contact and electric symmetry coefficients
as a way to measure pure energy currents.
\end{abstract}

\section{Introduction}

There is a growing interest in controlling and manipulating heat currents flowing in nanoscale devices~\cite{review}.
Efficient heat-to-work transformation, low-temperature thermometry, heat rectification and quantum cooling
are but a few examples of possible beneficial applications foreseen within the realm of mesoscopic conductors~\cite{review2}.
An equally important motivation is the study of the fundamental mechanisms that govern energy transfer
and dissipation beyond a purely (semi)classical description of particle transport.

In this work, we are concerned with heat fluxes driven by voltage biases. Current carriers (electrons)
also carry energy and represent the main contribution to heat in quantum electronics at low temperatures.
For very small shifts, the generated Peltier heat is reversible, its sign being dependent of the electric current direction. 
For larger shifts, heat is dominated by Joule power, which is always positive since it is connected to dissipation.
The measured heat current then consists of two terms (energy flux and Joule heating) and it is thus natural
to analyze their relative importance based on the transmission properties of the system. Here, we consider
a single-level quantum dot setup, a prototypical mesoscopic system that allows us to analyze the role
of strong electron-electron interactions in the generation of heat far from equilibrium.

The topic of voltage-driven quantum heat currents is interesting~\cite{review3} and has been examined in a number of
different systems~\cite{kul94,bog99,cip04,free06,zeb07,lei10,whi13,jia15,zim16}. Quantum dot setups
are specially appealing due to their experimental tunability and theoretical simplicity. 
Nonlinear heat current in quantum dots were addressed in Ref.~\cite{lop13,mea13}. It was argued that nonequilibrium screening
effects lead to charge buildups that affect the heat current--voltage characteristic beyond linear response.
The scattering model considered weak interactions and was therefore valid for large dots only. Coulomb blockade
effects were then analyzed by us in Refs.~\cite{sie14,sie15}. We found that Joule heating quickly surpasses the Peltier contribution
as the applied voltage increases. The problem was also investigated recently~\cite{ger15,yam15}.
We note in passing that the study of heat currents is strongly related to thermoelectrics
due to reciprocity. In linear response, the Seebeck and Peltier coefficients are connected
through the Kelvin-Onsager relation. Discussions beyond linear response are available
in the literature~\cite{sie15,iyo10,hwa13,bed13,cim14,sel14}.
The role of electron-electron interactions is crucial for the breakdown
of reciprocal relations when the driving fields are large~\cite{hwa13,mat14}.

Quantum dots with strong Coulomb interactions act as artificial quantum impurities.
In the limit of very low temperatures and strong coupling, the many-body spin interaction between the localized
electron in the quantum dot and the delocalized carriers in the attached reservoirs
leads to the Kondo effect, which can be detected via electric conductance measurements~\cite{exp1,exp2,exp3}.
It is worth mentioning that studies of the generated heat current in Kondo systems are scarce~\cite{boe576,sai13} and focusing only on the linear response regime.
Our aim here is to extend our work in Ref.~\cite{sie085} and investigate beyond linear response the dissipated power of a quantum dot in the Fermi liquid fixed point
where charge fluctuations are quenched and the Kondo singlet is well formed. This case is relevant for very low temperatures. We find
that the symmetry properties of the local density of states with respect to voltage
determine the energy flux and the dissipated power. Our results are relevant
for the evaluation of heat asymmetries that take place when voltage is reversed
or when heat is measured in different contacts, thus providing a better understanding
for the nonlinear electrothermal response of a Kondo-correlated system.

\section{Model Hamiltonian}
We consider a Kondo impurity in a quantum dot connected to two leads $\alpha = \{L,R\}$ characterized by their electrochemical potential $\mu_\alpha = \varepsilon_F +eV_\alpha$ and temperature $T_\alpha = T$. In the limit where interactions are infinite (charging energy $U\rightarrow\infty$), we consider the Anderson Hamiltonian in the slave-boson formalism~\cite{hewson}. The full Hamiltonian reads

\begin{eqnarray}
\mathcal{H}=\sum_{\alpha k \sigma} \varepsilon_{\alpha k} C^\dagger_{\alpha k \sigma} C_{\alpha k \sigma} + \sum_{\sigma} \varepsilon_d f^\dagger_{\sigma} f_{\sigma} + \sum_{\alpha k \sigma} \mathcal{V}_{\alpha k} \left(C^\dagger_{\alpha k \sigma} b^\dagger f_\sigma + \mathrm{H.c.}\right) + \lambda \left(b^\dagger b + \sum_\sigma f^\dagger_\sigma f_\sigma -1\right)\, .
\end{eqnarray} 

Here, the first term in the right-hand side represents the fermionic reservoirs. The operator $C^\dagger_{\alpha k \sigma}$ ($C_{\alpha k \sigma}$) creates (destroys) electrons in the $\alpha$ reservoir with energy $\varepsilon_{\alpha k}$, where $k$ and $\sigma$ are the wavenumber and the spin, respectively. The next term is the dot Hamiltonian. In this case, $f^\dagger_\sigma$ ($f_\sigma$) creates (annihilates) a pseudofermion with energy $\varepsilon_d$ and spin $\sigma$. The tunneling between the dot and the reservoirs is described in the next term. $\mathcal{V}_{\alpha k}$ is the amplitude of electrons hopping on and off the reservoir $\alpha$. Additionally, we add the auxiliary bosonic field $b^\dagger$, which creates an empty state in the dot. Finally, in order to ensure that the quantum dot is occupied with only one electron (we recall that $U\rightarrow\infty$)
we include a Lagrange multiplier $\lambda$ term in the Hamiltonian $\mathcal{H}$.
  


Let us consider the mean-field approach to the slave-boson Hamiltonian. This amounts to taking the leading order in a $1/N$ expansion ($N$ being the spin degeneracy). As a consequence, we replace the boson operator by its mean value $\langle b \rangle \equiv \tilde{b}$. This approximation neglects charge fluctuations since they are completely screened out. We remark that this approximation is valid in the Fermi liquid regime where temperatures are lower than the Kondo temperature $T_K$ and dot energy level lies below the Fermi energy $\varepsilon_F$. 

Within the nonequilibrium Green's function framework, we derive the mean-field equations for the expectation value of $b$ and the Lagrange multiplier $\lambda$. First, we compute the evolution of the boson operator in the stationary limit, $\sum_{\alpha k \sigma} \tilde{\mathcal{V}}_{\alpha k} G^<_{f\sigma, \alpha k \sigma}(t,t) =-iN\lambda |\tilde{b}|^2/\hbar$, where $\tilde{\mathcal{V}}_{\alpha k}=b_\alpha\mathcal{V}_{\alpha k}$ is the normalized tunneling amplitude through the dot and $G^<_{f\sigma,\alpha k \sigma}(t',t)=-(1/\hbar)\langle C^\dagger_{\alpha k \sigma}(t') f_\sigma(t)\rangle$ is the lesser Green's function for the tunneling process. The next expression corresponds to the single-occupied state condition in the Lagrange term. Its mean-field equation reads  $\sum_\sigma G^<_{f\sigma,f\sigma}(t,t) = i(1-N|b|^2)/\hbar$, where $G^<_{f\sigma,f\sigma}(t',t)=-(i/\hbar)\langle f_\sigma^\dagger(t') f_\sigma(t) \rangle$ is the dot pseudofermion lesser Green's function. Both mean-field equations can be combined in a complex self-consistent equation in the Fourier space,

\begin{eqnarray}
\frac{2}{\pi} \int_{-D}^D d\omega \frac{\mathcal{F}(\omega)}{\omega-\tilde{\varepsilon}_d+i\tilde{\Gamma}}=(\varepsilon_d-\tilde{\varepsilon}_d)\frac{N}{\Gamma}-i\left(1-N\frac{\tilde{\Gamma}}{\Gamma}\right) \, . \label{Eq:Selfeq}
\end{eqnarray}

The parameters $\tilde{\varepsilon}_d=\varepsilon_d+\lambda$ and $\tilde{\Gamma}=|\tilde{b}|^2 \Gamma$ are the renormalized level position and width of the Kondo resonance at the dot. $\Gamma=\Gamma_L+\Gamma_R$ is the hybridization of the energy level due to tunneling, where $\Gamma_\alpha = \pi \rho_\alpha(\omega) |\mathcal{V}_\alpha(\omega)|^2$ depends on the tunneling amplitudes and the density of states $\rho_\alpha(\omega)$ of the lead $\alpha$. In the wide band limit, $\rho$ is constant and nonzero inside the bandwidth $\omega<|D|$. Equation~(\ref{Eq:Selfeq}) also depends on the nonequilibrium distribution function $\mathcal{F}(\omega)=\sum_\alpha (\Gamma_\alpha/\Gamma) f_\alpha(\omega)$ which is, in this case, a weighted sum of the Fermi-Dirac distribution functions $f_\alpha(\omega)=1/\{1+\exp{[(\omega-\mu_\alpha)/(k_BT_\alpha)]}\}$ \cite{cab425}. 
In the slave-boson mean-field approach, the Kondo resonance arises when the auxiliary boson condenses and, as result, $\tilde{\Gamma}$
becomes $k_B T_K$ and $\lambda$ shifts $\varepsilon_d$ up to the Fermi level \cite{coleman}.

Once we solve Eq.~(\ref{Eq:Selfeq}) for both $\lambda$ and $|\tilde{b}|$, the transport properties can be readily analyzed.
We focus on the calculation of the heat current through the left reservoir $J\equiv J_L$, which is split into two different contributions: $J=J_E+J_{I}$.
The first term is the energy current flowing through the lead, $J_E=d(\sum_{k\sigma}  \varepsilon_{Lk} C_{Lk\sigma}^\dagger C_{Lk\sigma})/dt$.
The second term is the Joule heating, $J_{I}=-I_LV_L$, where $V_L$ is the voltage applied to the left reservoir and $I_L=-e d(\sum_{k\sigma} C_{Lk\sigma}^\dagger C_{Lk\sigma})/dt$ is the charge current given by the evolution of the left lead occupation.
Both charge and energy currents are conserved in the steady state
whilst the heat fluxes satisfy the relation $\sum_\alpha (J_\alpha+I_\alpha V_\alpha)=0$.

We find
\begin{eqnarray}
J_E&=&\frac{2}{h} \int_{-D}^D d\omega\, \mathcal{T}(\omega)(\omega-\varepsilon_F)[f_L(\omega)-f_R(\omega)]\, , \label{Eq:JE}\\
J_I&=&-\frac{2eV_L}{h} \int_{-D}^D d\omega\, \mathcal{T}(\omega)[f_L(\omega)-f_R(\omega)]\, .\label{Eq:JI}
\end{eqnarray}
Both currents depend on the Fermi functions of the leads and the transmission function, which takes a particularly simple form: $\mathcal{T}(\omega)=4\tilde{\Gamma}_L\tilde{\Gamma}_R/[(\omega-\tilde{\varepsilon}_d)^2+\tilde{\Gamma}^2]$.
It represents a Breit-Wigner lineshape with renormalized parameters ($\tilde{\varepsilon}_d$ and $\tilde{\Gamma}$).
Unlike the noninteracting case, however, the parameters of this transmission function depends on the applied voltages
and must be calculated from Eq.~(\ref{Eq:Selfeq}) for each dc bias.

\section{Results}

\begin{figure}
\centering
\includegraphics[scale=0.7]{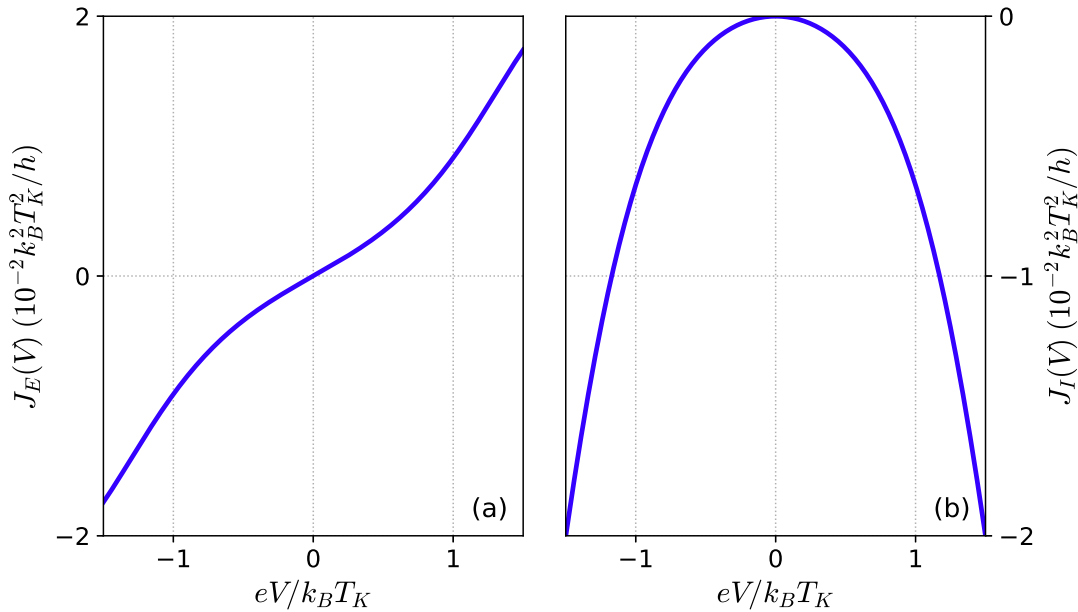}
\caption{(a) Energy current $J_E$ and (b) Joule dissipation current $J_I$ as a function of the applied voltage bias $V$ for a given value of the energy level of the dot ($\varepsilon_d=-2.5\Gamma$). Parameters: temperature $k_BT=0.01\Gamma$ and bandwidth $D=100\Gamma$. }
\label{fig:1}
\end{figure}

We investigate the energy and Joule currents given by Eqs.~\eqref{Eq:JE} and~\eqref{Eq:JI} in response to a symmetrically applied voltage bias ($\mu_L=eV/2$ and $\mu_R=-eV/2$ setting $\varepsilon_F=0$ as the reference energy). Hereafter, we assume symmetric tunnel couplings $\Gamma_L=\Gamma_R=\Gamma/2$.
In Fig~\ref{fig:1}, we depict $J_E$ and $J_I$ as a function of $V$ for a quantum dot energy level in the Kondo regime ($\varepsilon_d = -2.5\Gamma$). We focus on voltages smaller than the Kondo temperature $k_BT_K=D\exp{(-|\varepsilon_d|\pi/\Gamma)}$. For $V$ larger than $k_BT_K/e$
the boson field vanishes and the mean-field approach breaks down. For voltages smaller than $k_BT_K/e$ the boson is nonzero
and the results are thus reliable.

\begin{figure}
\centering
\includegraphics[scale=0.7]{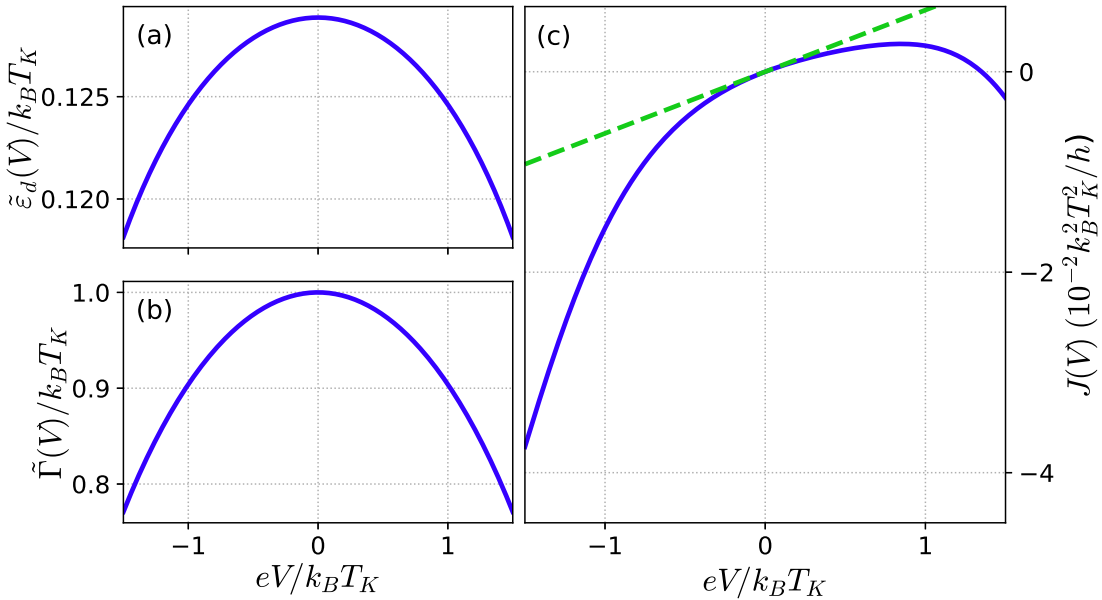}
\caption{(a) Renormalized level position $\tilde{\varepsilon}_d$ and (b) renormalized width $\tilde{\Gamma}$ of the Kondo resonance as a function of the voltage bias $V$ for values lower than the Kondo temperature. (c) Heat current $J$ as a function of the voltage bias $V$. The dashed green line exhibits the Peltier contribution to the heat current. The parameters are the same as in Fig.~\ref{fig:1}.}

\label{fig:2}
\end{figure}

The energy current in Fig.~\ref{fig:1}(a) is an increasing function of $V$
since for positive voltages the energy flows from the left lead and this energy
increases for higher values of $V$. For $V<0$ carriers predominantly impinge from the right lead
and $J_E$ decreases as $V$ becomes more negative. Clearly, the energy current 
shows an antisymmetric shape around $V=0$. In contrast,
the Joule term is always negative
(with our sign convention, dissipation is negative) and symmetric under the replacement $V\to -V$.
Let us examine in more detail the origin of these symmetry properties. First, we note that the difference of the Fermi functions in Eqs.~\eqref{Eq:JE} and~\eqref{Eq:JI}, $f_L-f_R$, is an odd function of $V$ whereas the transmission function is even $\mathcal{T}(\omega,V)=\mathcal{T}(\omega,-V)$ because the mean-field parameters that characterize the Kondo resonance are also even functions of the applied voltage [see Figs.~\ref{fig:2}(a) and~\ref{fig:2}(b)].
These findings are expected since the renormalized width should not depend on the direction of the voltage bias. Further,
the position of the Kondo peak lies near the Fermi level and is a function that weakly depends on $V$.
As a result of the combination of an odd and an even function, we infer that the energy current is antisymmetric when $V\rightarrow -V$. On the other hand, the Joule term $J_I(V)$ is symmetric due to the fact that there is an additional $V$ factor in front of the integral in Eq.~\eqref{Eq:JI}. 

The total heat current is shown in Fig.~\ref{fig:2}(c). At very low voltages, $J_E(V)$ dominates the heat transport over $J_I(V)$, showing a linear dependence [dashed line in Fig.~\ref{fig:2}(c)], which is the hallmark of the Peltier effect. Then, we can approximate $J(V)\simeq M_0 V$ with $M_0$ the electrothermal conductance~\cite{butcher}, directly connected to the thermopower via the Kelvin-Onsager relation. Now, the thermopower is nonzero only for asymmetric density of states. In our case, we find nonzero electrothermal conductances because the Kondo peak is not exactly located at the Fermi energy $\tilde{\varepsilon}_d\simeq\varepsilon_F$ [see Fig.~\ref{fig:2}(a)] due to the potential scattering term in the quantum dot. This results in a nonsymmetric transmission function $\mathcal{T}(\omega,V)\neq\mathcal{T}(-\omega,V)$. Additionally, we find a positive heat current at positive voltages indicating $M_0>0$. Therefore, the Kondo dot at low $V>0$ might serve as a cooler, although this property is quickly dominated by Joule heating at increasing voltages. As seen in Fig.~\ref{fig:2}(c), the Joule term overcomes the energy current inducing a negative heat flow. Importantly, $J(V)$ is not symmetric under the transformation $V\to -V$ because $J$ arises from the addition of symmetric and antisymmetric functions.

\section{Discussion}
The asymmetries shown in the calculated heat current can be analyzed in terms of contact and electric asymmetries, as suggested in Ref.~\cite{lee13} for molecular tunnel junctions.
The contact asymmetry $\Delta_C = J(V)-J(-V)$ measures the dissipated power under the reversal of the dc bias. The electric asymmetry $\Delta_E = J_L(V)-J_R(V)$ quantifies the different heat dissipation in both terminals as a function of $V$. Using the even and odd properties of $J_E(V)$ and $J_I(V)$ [Eqs.~(\ref{Eq:JE}) and (\ref{Eq:JI})], we straightforwardly arrive at the relation $\Delta_C(V) = 2J_E(V)$. This implies that the behavior of the contact asymmetry is in fact given by the results in Fig.~\ref{fig:1}a. Remarkably, we find that the electric asymmetry is also proportional to the energy current $\Delta_E(V) = 2J_E(V)$. This result is general~\cite{jar165} in the case of both a symmetric voltage bias $V_L=-V_R=V/2$ and a symmetric transmission function. Any quantum conductor that satisfies these conditions will show similar contact and electric asymmetries. Here, we find that these symmetry conditions are satisfied for an artifitial Kondo impurity due to the symmetric renormalization of the mean-field parameters. Therefore, the Kondo effect does not break these symmetries at least in the Fermi liquid regime. As a byproduct, our result would facilitate the experimental detection of the energy current from a measurement of either the heat current for different voltage directions or the heat current at different reservoirs for the same dc bias. 

\section{Conclusions}

To sum up, we have examined the heat flow through an artificial Kondo impurity connected to two leads with a symmetric applied voltage. Within the mean-field slave-boson formalism, we have obtained the expression for the transmission and have computed both the energy current and the Joule dissipation current. We find an antisymmetric energy current and a symmetric Joule term as a function of the dc bias. This behavior can be understood from the symmetry properties of the transmission function which arises from the symmetry of the renormalized width and energy position of the Kondo resonance. Additionally, we show that at voltages smaller than the Kondo temperature the Peltier effect first dominates the heat current while Joule heating governs the heat current at higher biases. Then, the crossover from linear to quadratic behavior is controlled by the Kondo temperature.
Our results could be tested with present techniques since there is a variety 
of available methods that measure heat currents in quantum conductors~\cite{lee13,mol92,chi06,mes09,jez13,cui17}.

We acknowledge support from MINECO under grant No.\ FIS2014-52564 and a PhD grant from CAIB.


\section*{References}
\begin{thebibliography}{9}
\bibitem{review}
Benenti G, Casati G, Saito K, and Whitney R S 2017 \textit{Phys. Rep.} \textbf{694} 1
\bibitem{review2}
Giazotto F, Heikkil\"a T T, Luukanen A , Savin A M and Pekola J P 2006
\textit{Rev. Mod. Phys.} \textbf{78}  217 
\bibitem{review3}
S\'anchez D and L\'opez R 2016 \textit{C. R. Physique}  \textbf{17} 1060 
\bibitem{kul94}
Kulik I O 1994 \textit{J. Phys.: Condens. Matter} \textbf{6} 9737
\bibitem{bog99}
Bogachek E N, Scherbakov A G, and Landman U 1999 \textit{Phys. Rev. B }\textbf{60} 11678
\bibitem{cip04}
\c{C}ipilo\v{g}lu M A, Turgut S, and Tomak M 2004
\textit{Phys. Stat. Sol. (b)} \textbf{241} 2575 
\bibitem{free06}
Freericks J K and Zlatic V, \textit{Condensed Matter Physics} \textbf{9} 603
\bibitem{zeb07}
Zeberjadi M, Esfarjani K, and Shakouri A 2007 \textit{Appl. Phys. Lett.} \textbf{91} 122104
\bibitem{lei10}
Leijnse M, Wegewijs M R, and Flensberg K 2010
\textit{Phys. Rev. B} \textbf{82} 045412
\bibitem{whi13}
Whitney R S 2013 \textit{Phys. Rev. B} \textbf{88} 064302
\bibitem{jia15}
Jiang J H, Kulkarni M, Segal D, and Imry Y 2015
\textit{Phys. Rev. B} \textbf{92} 045309 
\bibitem{zim16}
Zimbovskaya N 2016
\textit{J. Phys.: Condens. Matter} \textbf{28} 183002
\bibitem{lop13}
L\'opez R and S\'anchez D 2013 \textit{Phys. Rev. B} \textbf{88} 045129
\bibitem{mea13}
Meair J and Jacquod P 2013 \textit{J. Phys.: Condens. Matter} \textbf{25} 082201
\bibitem{sie14}
Sierra M A and S\'anchez D 2014 \textit{Phys. Rev. B} \textbf{90} 115313 
\bibitem{sie15}
Sierra M A and S\'anchez D 2015  \textit{Materials Today: Proceedings} \textbf{2} 483
\bibitem{ger15}
Gergs N M, H\"orig C B M,  Wegewijs M R, and Schuricht D 2015 \textit{Phys. Rev. B} \textbf{91} 201107(R)
\bibitem{yam15}
Yamamoto K and Hatano N 2015 \textit{Phys. Rev. E} \textbf{92} 042165
\bibitem{iyo10}
Iyoda E, Utsumi Y, and  Kato T 2010
\textit{J. Phys. Soc. Jpn.} \textbf{79} 045003
\bibitem{hwa13}
Hwang S Y, S\'anchez D, Lee M, and L\'opez R 2013
\textit{New J. Phys.} \textbf{15} 105012
\bibitem{bed13}
Bedkihal S, Bandyopadhyay M, and Segal D 2013
\textit{Eur. Phys. J. B} \textbf{86} 506 
\bibitem{cim14}
Cimmelli V A, Sellitto A, and Jou D 2014
\textit{Proc. Royal Soc. A} \textbf{470} 0265
\bibitem{sel14}
Sellitto A 2014 \textit{Physica D} \textbf{283} 56
\bibitem{mat14}
Matthews J, Battista F, S\'anchez D, Samuelsson P, and Linke H 2014
\textit{Phys. Rev. B} \textbf{90} 165428
\bibitem{exp1}
Goldhaber-Gordon D \textit{et al.} 1998 \textit{Nature} \textbf{391} 156 
\bibitem{exp2}
Cronenwett S M, Oosterkamp T H, and Kouwenhoven L P 1998 \textit{Science} \textbf{281} 540
\bibitem{exp3}
Schmid J, Weis J, Eberl K, and Klitzing K v 1998 \textit{Physica B} \textbf{256} 182
\bibitem{boe576} Boese D and Fazio R 2001 \textit{Europhys. Lett.} \textbf{56} 576
\bibitem{sai13}
Saito K and Kato T 2013 \textit{Phys. Rev. Lett.} \textbf{111} 214301
\bibitem{sie085} Sierra M A, L\'opez R and S\'anchez D 2017 \textit{Phys. Rev. B} \textbf{96} 085416
\bibitem{hewson}
Hewson A C 1997 \textit{The Kondo Problem to Heavy Fermions} (Cambridge University Press)
\bibitem{cab425}  Balseiro C A, Usaj G, and S\'anchez M J 2010 \textit{J. Phys.:
Condens. Matter} \textbf{22} 425602
\bibitem{coleman}
Coleman P 1984 \textit{Phys. Rev. B} \textbf{29} 3035
\bibitem{butcher}
Butcher P N 1990 \textit{J. Phys.: Condens. Matter} {\bf 2} 4869
\bibitem{lee13}
Lee W, Kim K, Jeong W, Zotti L A, Pauly F, Cuevas J C, and Reddy P 2013
\textit{Nature} \textbf{498} 209
\bibitem{jar165} Arg\"uello-Luengo J,  S\'anchez D, and L\'opez R 2015 \textit{Phys. Rev. B} \textbf{91} 165431

\bibitem{mol92}
Molenkamp L W, Gravier T, van Houten H, Buijk O J A, Mabesoone M A A,
and Foxon C T 1992 \textit{Phys. Rev. Lett.} \textbf{68} 3765
\bibitem{chi06}
Chiatti O,  Nicholls J T, Proskuryakov Y Y, Lumpkin N, Farrer I, and Ritchie D A 2006  \textit{Phys. Rev. Lett.} \textbf{97} 056601
\bibitem{mes09}
Meschke M, Guichard W, and Pekola J P 2009
\textit{Nature} \textbf{444} 187
\bibitem{jez13}
Jezouin S, Parmentier F D, Anthore A, Gennser U, Cavanna A, Jin Y, and Pierre F 2013
\textit{Science} \textbf{342} 601
\bibitem{cui17}
Cui L,  Jeong W, Hur S, Matt M, Kl\"ockner J C, Pauly F, Nielaba P,
Cuevas J C, Meyhofer E, and Reddy P 2017 \textit{Science} \textbf{355} 1192
\end{thebibliography}
\end{document}